\documentclass[10pt,twocolumn,aps,prl,showpacs,superscriptaddress,floatfix]{revtex4}

\usepackage{epsfig}
\usepackage{amsmath}
\usepackage{graphicx}
\usepackage{color}
\usepackage{hyperref}
\usepackage{bbm}
\usepackage{soul}

\newcommand{\ket}[1]{\vert{#1}\rangle} 
\newcommand{\bra}[1]{\langle{#1}\vert} 
 
\newcommand{\proj}[1]{\ket{#1}\!\bra{#1}}
\newcommand{\mean}[1]{\langle #1 \rangle}

\renewcommand\Re{\operatorname{Re }}
\renewcommand\Im{\operatorname{Im }}

\newcommand{\beq}{\begin{equation}}
\newcommand{\eeq}{\end{equation}}

\newcommand{\be}{\begin{equation}}
\newcommand{\ee}{\end{equation}}

\newcommand{\ben}{\begin{eqnarray}}
\newcommand{\een}{\end{eqnarray}}
\usepackage{amsmath}

\begin{document}

\title{ Contextuality in phase space}

\author{Ali Asadian}
\affiliation{{Naturwissenschaftlich-Technische Fakult\"at, Universit{\"a}t Siegen, Walter-Flex-Str.~3, D-57068 Siegen, Germany}}
\author{Costantino Budroni}
\affiliation{{Naturwissenschaftlich-Technische Fakult\"at, Universit{\"a}t Siegen, Walter-Flex-Str.~3, D-57068 Siegen, Germany}}
\author{Frank E. S. Steinhoff}
\affiliation{{Naturwissenschaftlich-Technische Fakult\"at, Universit{\"a}t Siegen, Walter-Flex-Str.~3, D-57068 Siegen, Germany}}
\author{Peter Rabl} 
\affiliation{Institute of Atomic and Subatomic Physics, TU Wien, Stadionallee 2, 1020 Wien, Austria}
\author{Otfried G\"uhne} 
\affiliation{{Naturwissenschaftlich-Technische Fakult\"at, Universit{\"a}t Siegen, Walter-Flex-Str.~3, D-57068 Siegen, Germany}}

\date{\today}

\begin{abstract}

We present a general framework for contextuality  tests in phase space using displacement operators. First, we derive a general condition that a single-mode displacement operator should fulfill in order to construct Peres-Mermin square and similar scenarios. This approach offers a straightforward scheme for experimental implementations of the tests via modular variable measurements. In addition to the continuous variable case, our condition can also be applied to finite-dimensional systems in discrete phase space,  using Heisenberg-Weyl operators. This approach, therefore, offers a unified picture of contextuality with a geometric flavor.

\end{abstract}

\pacs{ 07.10.Cm, 	
           03.65.Ta, 	
           03.65.Ud 	
           }
\maketitle

The concept of contextuality highlights the fundamental departure of the quantum description of the world from that of classical models.  In everyday life, the result of a measurement occurs irrespectively of which other compatible measurements (the so-called ``context") are simultaneously measured together with it. Quantum mechanics, however, as originally shown by Kochen and Specker \cite{Specker60,Kochen}, does not allow the noncontextual assignment of values to physical quantities, giving rise to testable differences in predictions of quantum and  noncontextual theories \cite{KCBS08}, as in Bell's theorem \cite{Bell64} (locality being a special case of noncontextuality). Remarkably, this can happen irrespectively of the quantum state, i.e., state-independent contextuality (SIC)\cite{Cabello08,BBCP09,YO2012,KBLGC12,RH14,CKB15}.
Contextuality has been confirmed in experiments with genuine microscopic systems, such as two qubits \cite{Roos09, NMR, Photons} and qutrits \cite{SciRep, Zeilinger,Zhang2013}, and it has been recognized as a critical resource for quantum computing \cite{HWVE14, Raussendorf13, DGBR14} and for several quantum information-theoretic tasks \cite{HHHHPB10,GBCKL14,CAEGCXL14}.
Recently, there has been a growing interest in exploring quantum features such as quantum superposition closer to the classical realm. Along this line, a far-reaching attempt, perhaps, is to experimentally probe the quantum contextuality in continuous variable systems with most ``classical-like" operations in phase space.

Phase space displacement operators exhibit favorable properties, making them very suitable 
for investigating contextuality in phase space. For one thing, their phase space 
functions given by Weyl-Wigner correspondence are bounded and, thus comply with the explicit 
assumption used for the derivation of the existing noncontextuality inequalities in which the 
classical variables take on a limited range of values. This is in contrast to, e.g., 
parity operators as they are described by an unbounded delta function in phase space. That is 
why parity correlation measurements can lead to the violation of Bell's inequality for states 
represented by positive Wigner functions \cite{Parity}. Moreover, displacement operators, unlike 
quantum operations such as Kerr nonlinearities, do not transform coherent states into 
nonclassical states characterized by negative-valued Wigner functions. Indeed, displacement 
operators reveal a characteristic geometrical feature of quantum mechanics in phase space 
originated from the canonical commutation relations between position and momentum operators.

In this Letter, we present a general approach to quantum contextuality in phase space, and in particular to SIC. This is done by giving an explicit recipe to witness SIC in continuous variable systems. Our results include as a special case a previous proposal of observing quantum contextuality in infinite dimensions  \cite{PlastinoPRA2010}.
Our approach offers a simple and optimal scheme for experimental implementations of the tests via measurements of modular variables, i.e., the Hermitian components of the displacement operator. Modular variables play a significant role in various tests of quantum nonclassical effects, such as the GHZ theorem \cite{Massar}, macroscopic realism \cite{Asadian} and entanglement detection \cite{Hornberger, Zhang}.

\emph{Preliminary notions.---} 
Let us consider nine dichotomic observables $A_{jk}$, $j,k=1,2,3$, such that triplets of  observables sharing  a common index $j$ or $k$ are compatible. Here, compatibility means that they can be measured jointly or sequentially without disturbing each other. In quantum mechanics this is, for instance, the case if the observables commute. Thus, we can measure the following mean values \cite{Cabello08}:
\begin{align}
\label{pmoperator}
\mean{\tilde\chi_{\rm PM}} = & \mean{A_{11}A_{12}A_{13}}+ \mean{A_{21}A_{22}A_{23}} +
\mean{A_{31}A_{32}A_{33}}  \nonumber \\ & +\mean{A_{11}A_{21}A_{31}} + \mean{A_{12}A_{22}A_{32}}
- \mean{A_{13}A_{23}A_{33}}.
\end{align}
Such conditions are satisfied by the following choice of 
$\{A_{jk}\}$, also known as Peres-Mermin (PM) square~\cite{Peres90,Mermin90},
\begin{equation}\label{mpsquare}
\begin{matrix} 
A_{jk} & k=1 & k=2 &   k=3\\
\hline
j=1 &\sigma_z \otimes \openone&    \openone \otimes \sigma_z &   \sigma_z \otimes \sigma_z\\
j=2 & \openone \otimes \sigma_x &    \sigma_x \otimes \openone &   \sigma_x \otimes \sigma_x\\
 j=3 &   \sigma_z\otimes\sigma_x  &  \sigma_x \otimes\sigma_z &  \sigma_y \otimes\sigma_y
 \end{matrix}
\end{equation}
leading  to a value of $\mean{\tilde\chi_{\rm PM}} = 6$, for any quantum state, since the product of operators along each row and column is $\openone$, except in the last column, where it is $-\openone$.
On the other hand, the maximal value of $\mean{\tilde\chi_{\rm PM}}$ is 4 for all noncontextual hidden variable (NCHV) theories, i.e., classical probability theories assigning a definite  $\pm 1$ value to each observable independently of the measurement context. This can be verified by a direct substitution of all possible $\pm1$ values. Hence, the PM square exhibits SIC.

To obtain the same construction in phase space, we introduce the single-mode displacement operator, $\mathcal{D}(\alpha)=e^{\alpha a^\dag-\alpha^* a}$,
where $a^\dag$ ($a$) is the creation (annihilation) operator of a single bosonic mode and 
$\alpha$ is a complex displacement amplitude. By using the Baker-Campbell-Hausdorff formula, i.e., $e^X e^Y=e^{X+Y}e^{[X,Y]/2}$, where $[X,Y]$ is a constant, we obtain the known relations
\begin{equation}
\label{eq:prodrule}
\mathcal{D}(\alpha_i)\mathcal{D}(\alpha_j)=e^{i\Im\{\alpha_i\alpha_j^*\}}\mathcal{D}(\alpha_i+\alpha_j),
\end{equation}
and
\begin{equation}
\label{eq:BCHcommut}
\mathcal{D}(\alpha_i)\mathcal{D}(\alpha_j)=e^{2i\Im\{\alpha_i\alpha_j^*\}} \mathcal{D}(\alpha_j) \mathcal{D}(\alpha_i).
\end{equation}
The phase factor indicates that displacements in different directions do not commute in general. 
\begin{figure}[t]
\begin{center}
\includegraphics[width=0.6\columnwidth]{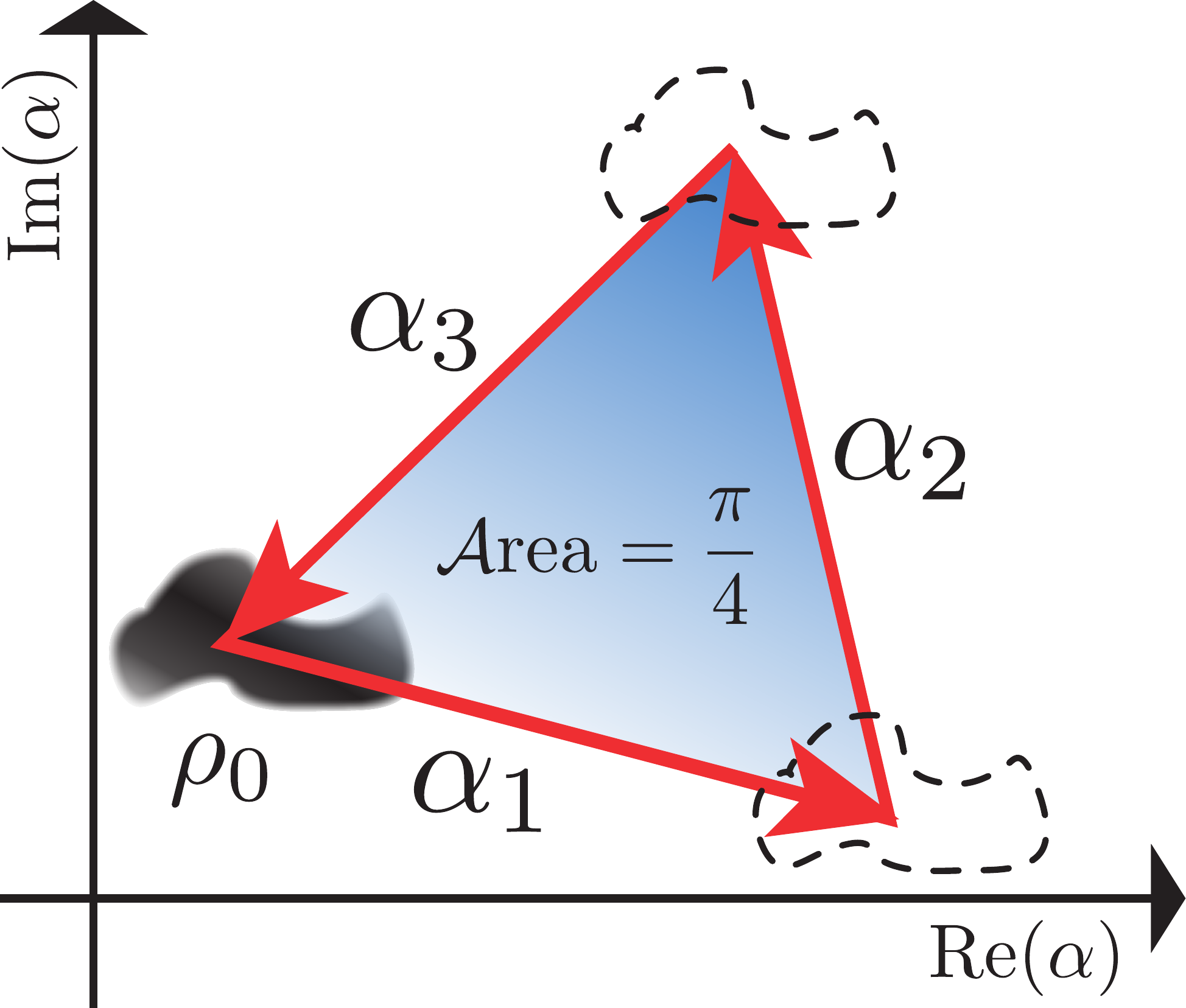}
\end{center}
\caption{ Pictorial representation of three displacement amplitudes $\alpha_i$ which span a triangle of area $\mathcal{A}=\pi/4$ and therefore satisfy the condition in Eq.~\eqref{eq:condition} used for constructing a PM square in phase space.}
\label{fig:area}
\end{figure}

\emph{Phase-space PM Square.---}
Geometrically, it is illuminating to think of the amplitudes $\alpha$ as real vectors in a two-dimensional space, i.e., $\alpha:=(\Re\alpha,\ \Im\alpha )$, which gives
\begin{equation}
\Im\{\alpha_i\alpha_j^*\}=\alpha_i\times\alpha_j:=|\alpha_i||\alpha_j|\sin\theta_{ij},
\end{equation}
 with $\theta_{ij}$ the angle between the two vectors.  Now, consider a set of amplitudes $\alpha_1, \alpha_2,$ and $\alpha_3$ satisfying the constraint
\begin{equation}
\label{eq:condition}
\alpha_1\times\alpha_2=\alpha_2\times\alpha_3=\alpha_3\times\alpha_1=\pm\dfrac{\pi}{2},
\end{equation}
from which $\alpha_1+\alpha_2+\alpha_3=0$ follows. This constraint yields the following algebraic relations for the displacement operators
similar to Pauli matrices, namely, 
\begin{equation}
\label{eq:Pauli}
\begin{split}
\mathcal{D}(\alpha_1)\mathcal{D}(\alpha_2)=\pm i \mathcal{D}(-\alpha_3)  \ , \ 
\{\mathcal{D}(\alpha_i), \mathcal{D}(\alpha_j)\}=0.
\end{split}
\end{equation}
A simple example satisfying condition \eqref{eq:condition} is given by amplitudes $\alpha_i$ with equal lengths $|\alpha_i|=\sqrt{\pi/\sqrt{3}}\simeq 1.34$ and rotated by an angle $2\pi/3$ with respect to each other.
As we will show below, this condition also extends to the discrete phase space.

 From Eq. \eqref{eq:Pauli} we can straightforwardly construct a PM square analogous to \eqref{mpsquare} for two bosonic modes,
\begin{equation}
\label{PMps}
\begin{matrix} 
A_{jk} & k=1 & k=2 &   k=3\\
\hline
j=1 &\mathcal{D}_{1}(-\alpha_1)&    \mathcal{D}_{2}(-\alpha_1) &   \mathcal{D}_{1}(\alpha_1) \mathcal{D}_{2}(\alpha_1)\\
j=2 & \mathcal{D}_{2}(-\alpha_2) &     \mathcal{D}_{1}(-\alpha_2) &   \mathcal{D}_{1}(\alpha_2)\mathcal{D}_{2}(\alpha_2)\\
 j=3 &   \mathcal{D}_{1}(\alpha_1) \mathcal{D}_{2}(\alpha_2) &  \mathcal{D}_{1}(\alpha_2) \mathcal{D}_{2}(\alpha_1)&  \mathcal{D}_{1}(\alpha_3)\mathcal{D}_{2}(\alpha_3)
 \end{matrix}
\end{equation}
where $\mathcal{D}_1$($\mathcal{D}_2$)  denotes a displacement operator for mode 1(2).
Operators $A_{jk}=A_{jk}^R+iA_{jk}^I$ within each row or column are mutually commuting, and the same holds for their real and imaginary Hermitian parts, i.e., the modular variables $A_{jk}^R$ and $A_{jk}^I$. The product of three operators in each row and 
column is $\mathbbm{1}$, except in the last column where it is  $-\mathbbm{1}$. The minus sign arises as a geometric phase proportional to the area $\mathcal{A}=\pi/4$ covered by the corresponding loops in phase space, cf.~ Fig. \ref{fig:area}. 
 
The displacement operators in \eqref{PMps} do not represent physical observables, and in each experimental run, we can only measure either the real or imaginary parts, $A_{jk}^R$ and $A_{jk}^I$. Therefore, instead of Eq.~\eqref{pmoperator}, we now consider the real-valued form
\begin{equation}
\label{eq:chiCV}
\mean{\chi_{PM}}=\mean{R_1}+\mean{R_2}+\mean{R_3}+\mean{C_1}+\mean{C_2}-\mean{C_3},
\end{equation}
where $R_j=\text{Re}(A_{j1}A_{j2}A_{j3})$ for products of three displacement operators in the same row and, analogously, $C_k=\text{Re}(A_{1k}A_{2k}A_{3k})$ for operator products in the same column. Explicitly, in terms of modular operators we obtain
\begin{equation}\label{eq:R}
\begin{split}
R_j=(A_{j1}^R A_{j2}^R - A_{j1}^I A_{j2}^I)A_{j3}^R - (A_{j1}^I A_{j2}^R+A_{j1}^R A_{j2}^I)A_{j3}^I,\\
C_k=(A_{1k}^R A_{2k}^R - A_{1k}^I A_{2k}^I)A_{3k}^R - (A_{1k}^I A_{2k}^R+A_{1k}^R A_{2k}^I)A_{3k}^I.
\end{split}
\end{equation} 

{\it Classical bound.---} Since only $A_{jk}^R$ or $A_{jk}^I$ can be measured, and not directly $A_{jk}$, we must compare the observed values with the most general NCHV theory, where those observables are associated with independent variables. In this case, the products $R_j$ and $C_k$ can assume values between $-2$ and $+2$, and for correlations measured as in Eqs. (\ref{eq:R}), we obtain an upper bound of $12$, which is the double of the quantum value $\mean{\chi_{\rm PM}}_{\rm QM}=6$.
To recover a lower classical bound, we enforce the additional constraints 
$\mean{(A_{jk}^R)^2+(A_{jk}^I)^2}\leq 1$, satisfied in QM, to our NCHV model,
by adding ``punishment'' terms that lower the classical bound whenever such a condition is 
violated (see, e.g., \cite{Guhne2010,YO2012, Guhne13}). Thus, we can define
\begin{equation}\label{PMpun}
\chi_{\rm PM}^{\rm pun} =\chi_{\rm PM} - \lambda \sum_{jk} \left| (A_{jk}^R)^2+ (A_{jk}^I)^2 -1\right|
\end{equation} 
and proceed to maximize the expectation value of $\chi_{\rm PM}^{\rm pun}$ with the only assumption that $A_{jk}^R$ and $A_{jk}^I$ are represented by noncontextual classical random variables $a_{jk}^R,a_{jk}^I$ taking values in $[-1,1]$. 
It can be proven (cf. Appendix) that for $\lambda \geq 2$, the maximal value for the expression  \eqref{PMpun} in any NCHV theory is $3\sqrt{3}$. This extends the 
bound derived by Plastino and Cabello~\cite{PlastinoPRA2010}, which holds only for a restricted set of NCHV theories where the condition $\mean{(A_{jk}^R)^2+(A_{jk}^I)^2}\leq 1$ is assumed, and makes contextuality tests in phase space generally applicable. 

\emph{Contextuality with a single mode.---}
\newcommand{\DD}{\mathcal{D}}
In close analogy to the case of qubits, the PM square~\eqref{PMps} has been constructed for two modes. A natural question arises, whether a similar set of operators with identical relations could already be identified for a single continuous-variable degree of freedom.
As we show now, this is not possible in a strict sense, but a PM square for single mode displacement operators can be realized with an arbitrary good approximation, when large displacements are allowed. 

Let us consider a set of nine single mode displacement operators $A_{jk}=\DD(\vec{A}_{jk})$, 
labeled by the corresponding two-dimensional displacement vectors $\vec{A}_{jk}$. To obtain the 
same relations as for the two mode case, we require first of all, that the product of three 
operators within each row and column commute and are proportional to $\openone$. This is given by the conditions 
 $\vec{A}_{j1}\times \vec{A}_{j2}= \vec{A}_{j2}\times \vec{A}_{j3}=\vec{A}_{j3}\times \vec{A}_{j1}=k\pi$, etc, for some integer $k$, implying $\vec{A}_{j1}+\vec{A}_{j2}+\vec{A}_{j3}=0$, etc.
Finally, the product of all three displacements is $+\openone$ meaning that $k$ must be even,  except for the third column where the product is $-\openone$, and therefore, $k$ must be odd.  
By putting together the above relations for the displacement vectors we prove the following result:

\noindent
{\bf Observation.} It is not possible to represent the commutation
and product relations of the PM square with displacement
operators on a single mode. 

By the linear relations, it is sufficient to consider four vectors $\vec{A}_{11}, \vec{A}_{12}, \vec{A}_{21}, \vec{A}_{22}$, and define $\vec{A}_{13}= -\vec{A}_{11} - \vec{A}_{12}$ etc. Then, the cross product relations give us
\begin{align*}
\vec{A}_{11} \times \vec{A}_{12} = k_1 \pi , \; 
\vec{A}_{21} \times \vec{A}_{22} = k_2 \pi , \;
\vec{A}_{31} \times \vec{A}_{32} = k_3 \pi,
\\
\vec{A}_{11}\times \vec{A}_{21} = k_4 \pi , \; 
\vec{A}_{12} \times\vec{A}_{22} = k_5 \pi , \;
\vec{A}_{13}\times\vec{A}_{23}  = k_6 \pi,
\end{align*}
where $k_1, ... , k_5$ are even, while $k_6$ is odd. A direct, but tedious, calculation allows us to remove the vector variables, obtaining a conditions only for the integers $k_i$
\begin{align*}
(k_1-k_2)^2 + (k_1-k_3)^2 + (k_2-k_3)^2 - k_1^2- k_2^2-k_3^2
=
\\
(k_4-k_5)^2 + (k_4-k_6)^2 + (k_5-k_6)^2 - k_4^2- k_5^2-k_6^2.
\end{align*}
From this, one can directly see that, if the first five
$k_i$ are even, the remaining $k_6$ has to be even. The other 
interesting cases, where, e.g. three $k_i$ are even and the three remaining 
$k_j$ are odd, are also not possible. This proves the claim.
\begin{figure}[t]
\begin{center}
\includegraphics[width=0.4\columnwidth]{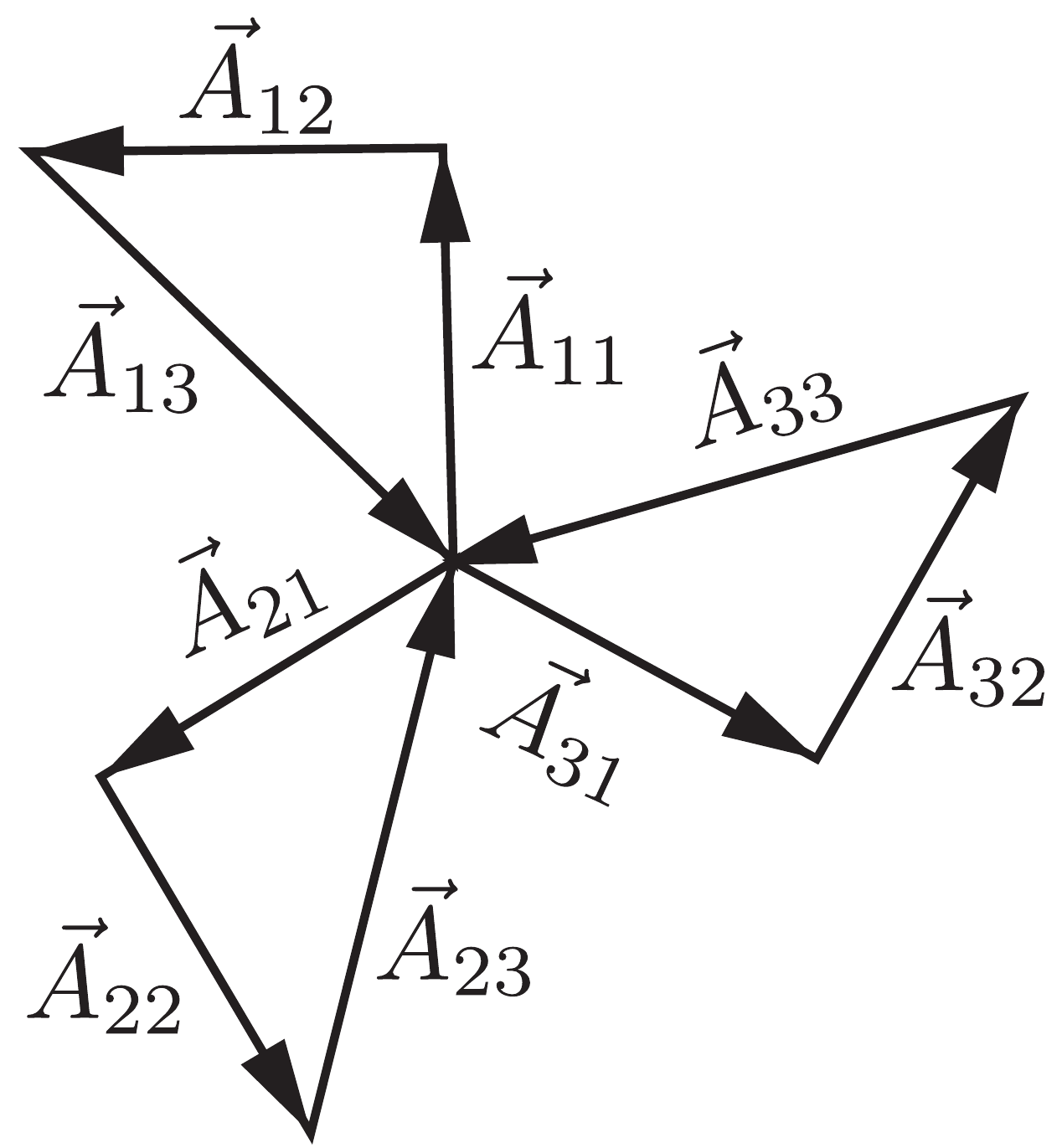}
\end{center}
\caption{Illustration of the displacement amplitudes $\vec A_{jk}$ used for the approximate construction of the PM square for a single mode. 
See the text for further details.}
\label{mpsingle}
\end{figure}

Nevertheless, as we will show now, it is possible 
to approximate the PM square on a single mode arbitrarily 
well, if the displacement is allowed to be large. Since any 
experimental implementation of the perfect PM square 
is only an approximation, this can still be a way to test hidden 
variable models, if the occurring errors are quantified
by additional measurements of the compatibility \cite{Guhne2010}.

To start, we set the vectors of the first row as 
\begin{align}
\vec{A}_{11}={{0}\choose{\ell}}, \;\;\;
\vec{A}_{12}={{-\ell}\choose{0}}, \;\;\;
\vec{A}_{13}={{\ell}\choose{-\ell}},
\end{align}
implying that the length of $\vec{A}_{13}$ is $L= \sqrt{2} \ell$.
The other rows are defined in a similar way, but rotated by $2\pi/3$ (cf. 
Fig.~\ref{mpsingle}).
Again, we have to impose the conditions on the area of the spanned triangles, but here we make a slightly different choice, by demanding that the product of five rows or columns should
be $- \openone$, while for the remaining one it should be $+\openone.$ One can easily see that this leads to the same contradiction as the usual PM square.

Our conditions can then be written as follows:
\begin{align}
\frac{\ell^2}{2}= k_i \frac{\pi}{2}
\label{cond1}
\end{align}
with odd $k_i$ for $i=1,2,3$, corresponding to the three rows.
For the first and second column, it follows that 
\begin{align}
\frac{\ell^2}{2}\sqrt{\frac{3}{4}}= k_i \frac{\pi}{2},
\label{cond2}
\end{align}
with odd $k_i$ for $i=4,5$. The factor $\sqrt{3/4}$ comes from the fact
that the vectors form an equilateral triangle (see Fig.~\ref{mpsingle}). Finally, for the last column,
we have
$
\tfrac{L^2}{2}\sqrt{\tfrac{3}{4}}= k_6 \tfrac{\pi}{2}
$
with an even $k_6.$ This, however, follows automatically from the previous
condition, since $L^2=2 \ell^2$. With odd $r$ and $s$ big enough, we can approximate arbitrarily well
 $\sqrt{3/4}$, i.e.,
\begin{equation}
\frac{r}{s} \approx \sqrt{\frac{3}{4}}.
\end{equation}
We then choose $\ell$ as $\ell = \sqrt{s \pi}$, and the conditions in Eq.~(\ref{cond1}) are automatically fulfilled. Furthermore, with this choice we have
\begin{equation}
\frac{\ell^2}{2}\sqrt{\frac{3}{4}}= s \sqrt{\frac{3}{4}} \frac{\pi}{2}
\approx r \frac{\pi}{2}
\end{equation}
so the conditions from Eq.~(\ref{cond2}) are approximately fulfilled. 
The approximation improves with increasing values of $r$ and $s$, which
requires an increasing displacement $\ell$. To give a concrete example, 
with displacements up to a maximal $|\alpha|=10$, the condition is $L=\sqrt{2 \pi s}< 10$
allows for a value of $s=15$. Then, the choice $r=13$ leads
to an approximation of Eq.~(\ref{cond2}) with a value $k_i=12.990$
instead of the ideal $k_i=13$. This is already a very good approximation
of the PM square.

\emph{Discrete phase space.---} 
The generality of our approach also allows us to investigate contextuality also in discrete phase space, i.e., a particle taking on discrete position values with a finite number of position sites. The single-step momentum and position displacements in a $d$-dimensional Hilbert space $\mathcal{H}$, respectively, are \cite{Vourdas2004},
\begin{equation}
\label{eq:HW}
Z=e^{i2\pi Q/d}, \ \ \ \ \ \ X=e^{-i2\pi P/d}
\end{equation}
with $Q=\sum_{n=0}^{d-1} n\proj{n} $ and ${P=\sum_{k=0}^{d-1} k\proj{k}}$,
the discrete position and momentum operators, respectively. The operators \eqref{eq:HW} are known as Heisenberg-Weyl (HW), or generalized Pauli operators.
The position and momentum basis states are related via discrete Fourier transform
\begin{equation}
\ket{n}=\dfrac{1}{\sqrt{d}}\sum_{k=0}^{d-1}e^{-i2\pi n k}\ket{k}, \ \ \ \ket{{k}}=\dfrac{1}{\sqrt{d}}\sum_{n=0}^{d-1}e^{i2\pi n k}\ket{n},
\end{equation} 
where the sum is defined modulo $d$, and they obey $X^m\ket{n}=\ket{n+m}$ and ${Z^l\ket{k}=\ket{k+l}}$. The periodic boundary condition gives $Z^d=\mathbbm{1}$ and $X^d=\mathbbm{1}$.
The commutation relation between $Z$ and $X$ operators are 
\begin{equation}
\label{UVcomm}
Z^lX^m=X^mZ^l e^{i2\pi l m/d},
\end{equation}
and a general displacement operator can be defined  as
\begin{equation}
\label{eq:DisDisplace}
\mathcal{D}(l,m):=Z^lX^m e^{-i\pi lm/d},
\end{equation}
describing the displacement of position and momentum with $m$ and $l$ steps, respectively.
With the shorthand notation $\alpha_j=\sqrt{\pi/d}(l_j-im_j)$, then, Eqs. \eqref{UVcomm} and \eqref{eq:DisDisplace} give a product rule for HW operators analogous to Eqs. \eqref{eq:prodrule} and \eqref{eq:Pauli}.  The condition established in the continuous limit, i.e., Eq. \eqref{eq:condition}, for the discrete scenario then becomes
\begin{equation}\label{eq:ConditionDiscrete}
m_2l_1-m_1l_2=m_3l_2-m_2l_3=m_1l_3-m_3l_1=d/2.
\end{equation} 
Since the left hand side of this equation  is always an integer, the dimension $d$ must be an even number. As an example, for $d=2$, the hopping steps are $m_1=0,  \:  l_1=1,  m_2=1, \:  l_2=0, \:  m_3=1,  \:  l_3=1$ giving $\mathcal{D}(\alpha_1)=\sigma_z$, $\mathcal{D}(\alpha_2)=\sigma_x$ and $\mathcal{D}(\alpha_3)=-i\sigma_z\sigma_x=\sigma_y$, respectively, and thus the PM square \eqref{mpsquare} for two-qubit systems is recovered. 
\begin{figure}[t!]
\centering
\includegraphics[width=1\linewidth]{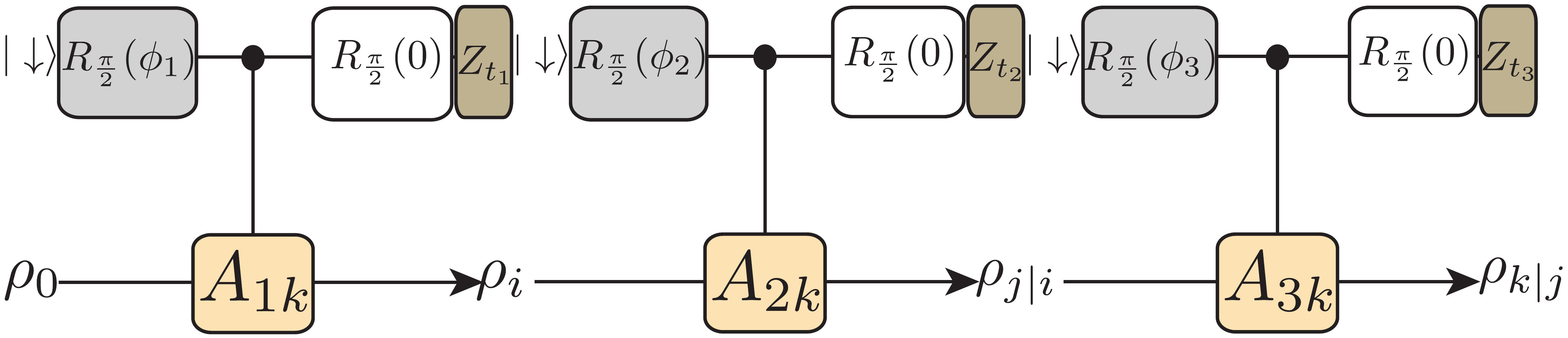}
\caption{A quantum circuit composed of a sequence of three Ramsey measurements used for measuring the correlations between modular variables.}
\label{fig:Sequence}
\end{figure} 

\emph{Implementation}.--- 
The experimental violation of the classical bound for $\chi_{\rm PM}^{\rm pun}$ requires the correlation measurements of modular operators appearing in \eqref{eq:R}.  By adapting techniques described in Refs. \cite{Asadian, Zhang2006}, this can be achieved in current trapped ion experiments and, eventually  with macroscopic mechanical resonators also. In general, we consider a qubit-oscillator model whose Hamiltonian reads ($\hbar=1$)
\begin{equation}
\label{eq:Htot}
H=  \omega_0\proj{\uparrow} +\omega b^\dag b + \Big [\lambda(t) b^\dag+\lambda^*(t)b\Big]\proj{\uparrow},
\end{equation}
where $\omega_0$ is the qubit splitting and the $\lambda(t)$ are time-dependent couplings which, in trapped ions, for example, can be realized by state-dependent optical forces (see, e.g., \cite{Monroe96, Monroe05}). The mode $b$ is the linear combination of two (spatial) orthogonal modes: $b=\cos(\theta)a_1+\sin(\theta) a_2$.

Now, we consider the sequence of three Ramsey measurements shown in Fig. (\ref{fig:Sequence}). At the beginning of each sequence $i=1,2,3$, the qubit is first prepared in a superposition $(|\downarrow \rangle + e^{i\phi_i}|\uparrow\rangle)/\sqrt{2}$ by applying a fast $\pi/2$ rotation $R_{\frac{\pi}{2}}(\phi_i)$. In a frame rotating with the bare qubit frequency  $\omega_0$ the coupled qubit-oscillator system then evolves under the action of $H$ for a time $\tau_i$
\begin{equation}
U(\tau_i)=\Big[\proj{\downarrow}+e^{i\phi(\tau_i)}\mathcal{D}_b[\alpha(\tau_i)]\proj{\uparrow}\Big]U_0(\tau_i),
\end{equation}
with $U_0(\tau_i)=e^{-i\tau_i\omega  b^\dag b}$ being the oscillator's free evolution.  In addition, the system accumulates phase $\phi(\tau_i)$.
Finally, the qubit is rotated back by applying $R_{\pi/2}(0)$, the population difference $Z$ is measured and the qubit is reset into the state $|\downarrow\rangle$. As shown in more detail in \cite{Asadian}, the average outcome of this measurement is 
\begin{equation}
\langle Z_{t_i}\rangle =  \langle Q(\varphi_i,\alpha_{t_i}) \rangle = \langle \cos \left( \varphi_i + \alpha_{t_i} b^\dag + \alpha_{t_i}^* b\right)\rangle,
\end{equation}
where $\alpha_{t_i}\equiv \alpha(\tau_i)e^{i\omega t_i}$ and $\varphi_i=\phi_i+\phi(\tau_i)$. Therefore, by choosing either $\varphi_i=0$ or $\varphi_i=\pi/2$ and by appropriately adjusting $\alpha_i$ and $\theta$ (for example, by changing the laser direction), we can measure $\langle A_{jk}^R\rangle$ and $\langle A_{jk}^I\rangle$. Further, in the case of interest where different  displacement operators commute, it can be shown \cite{Asadian} that  
 \begin{equation}
\label{eq:3point}
\mean{Z_{t_1}Z_{t_2}Z_{t_3}}=\text{Tr}\{Q(\varphi_1, \alpha_{t_1})Q(\varphi_2, \alpha_{t_2})Q(\varphi_3, \alpha_{t_3})\rho_0\},
\end{equation}
which then allows one to measure the operator products $R_{k}$ and $C_{j}$ given in Eqs. \eqref{eq:R} in terms of correlations between the outcomes of three Ramsey measurements. In summary all terms appearing in the expression for $\chi_{\rm PM}^{\rm pun}$ can be measured in this way. The symmetric choice of single-mode amplitudes given below Eq. \eqref{eq:Pauli} requires displacements of only $|\alpha_i|\simeq1.34$, which is well within what is achievable in current ion trap experiments. As an explicit example we provide a detailed description of the implementation of $C_3$ in the appendix.

\emph{Conclusions and outlook.---} 
The geometric feature of quantum mechanics demonstrated by phase space displacements leads to the impossibility of a noncontextual assignment of measurement results. We derived a general condition, which is sufficient for constructing various SIC scenarios both in continuous as well as in discrete phase space. We showed that displacement (or Heisenberg-Weyl) operators offer a unified geometric picture of contextuality. Moreover, our approach presents a simple and symmetric choice of modular variables,  simplifying, considerably, the experimental implementations using the state-of-art techniques of trapped ions.

In light of this, it would be interesting to develop our approach to Bell nonlocality tests, and construct Bell-like inequalities for probing nonlocality with suitable single-mode displacement operators. 
Finally, it can be shown that the geometric phase $\mathcal{A}$ giving rise to contextuality diminishes under decoherence and disappears in classical limit. This motivates us to consider it as a quantum resource which can be used for characterizing and quantifying contextuality.

\emph{Acknowledgments.} The authors thank C. Brukner and G. J. Milburn for stimulating discussions.   This work was supported by the EU (Marie Curie Grant No. CIG 
293993/ENFOQI, Project SIQS), the BMBF (Chist-Era Project QUASAR), the FQXi Fund (Silicon Valley 
Community Foundation),  the program ``Science without Borders" from brazilian agency CAPES, the DFG and the Austrian Science Fund (FWF) through SFB FOQUS, the Erwin
Schr\"odinger Stipendium No. J3653-N27, and the START Grant Y 591-N16.

\begin{center}
{\bf APPENDIX}
\end{center}
\vspace{0.5cm}
The appendix provides detailed derivations of the classical bound and the physical implementation of the measurements involved in the tests.

\section{I. Explicit computation of the classical bound}

 To simplify the notation, let us denote the classical variable associated with $A_{jk}^R$ and $A_{jk}^I$, respectively, as $x_{3k+j}$ and $y_{3k+j}$, and define the vector $\mathbf{X}=(x_1,\ldots,x_9,y_1,\ldots,y_9)$. Eq. \eqref{PMpun} can then be rewritten as 
\begin{equation}
\chi_{PM}^{pun}(\mathbf{X})= F(\mathbf{X})-\lambda P(\mathbf{X}),
\end{equation}
where $F$ is a function giving the PM expression of Eq. \eqref{eq:chiCV} and $P$ the function giving the punishment term. The classical bound is given by $\max_\mathcal{S} \chi_{PM}^{pun}$, where ${\mathcal{S}=\{ |x_i|\leq 1, |y_i|\leq 1\}}$. We then split our domain as $\mathcal{S}=\mathcal{B}\cup \mathcal{R}$, where 
\begin{equation}
\mathcal{B}=\{ |x_i^2+y_i^2| \leq 1\}, \qquad \mathcal{R}=\{ |x_i^2+y_i^2| > 1\}\cap \mathcal{S}.
\end{equation}
We can easily compute $\max_\mathcal{B} \chi_{PM}^{pun}$ as follows. Since $F$ is a harmonic function, i.e., $\Delta F=0$, and $\mathcal{B}$ is a compact set, i.e. closed and bounded, the maximum of $F$ is achieved on the border $\partial\mathcal{B}$. On the other hand, the minimum of $P$ is clearly achieved on $\partial\mathcal{B}$ and it is $0$. Such a value for $F$ has been computed in Ref.~\cite{PlastinoPRA2010}.

To complete our proof, we need to show that for appropriate values of $\lambda$, this is also the maximum of $\chi_{PM}^{pun}$ on $\mathcal{S}$. First, we notice that we can get rid of the absolute value in $P$, i.e. $P(\mathbf{X})_{|_\mathcal{R}}=\|X\|^2 - 9$. We can then compute the partial derivative of $\chi_{PM}^{pun}(\mathbf{X})$ along the direction $\mathbf{X}_{i}=(0,\ldots,0,x_i,0,\ldots,0,y_i,0,\ldots,0)$., i.e. $\nabla\chi_{PM}^{pun}(\mathbf{X}) \cdot \mathbf{X}_{i}/\|\mathbf{X}_{i}\|$. 
A simple calculation shows that
\begin{equation}
\nabla\chi_{PM}^{pun}(\mathbf{X}) \cdot \mathbf{X}_{i}/\|\mathbf{X}_{i}\|= \frac{R_i + C_i}{\sqrt{x_i^2+y_i^2}}- 2 \lambda \frac{x_i^2+y_i^2}{\sqrt{x_i^2+y_i^2}},
\end{equation}
which is always non-positive for 
\begin{equation}
\lambda \geq \frac{\max(R_i+C_i)}{2\min(x_i^2+y_i^2)}=\frac{4}{2}=2.
\end{equation}

Notice that $R_i+C_i$, as well as $F$, can be maximized on the cube $\mathcal{S}$ by simply taking the maximum over all values on the vertices of the cube, since it is a harmonic function.

The last thing to prove is that every point in the cube can be reached from a point in $\mathcal{B}$ with finite increments of $\mu_i\widetilde{\mathbf{X}}_{i}$, i.e., by moving in directions where partial derivative is non-positive. This follows from the fact that for all $\mathbf{X}\in \mathcal{S}$, there exist $\mathbf{X}_0\in \mathcal{B}$ and numbers $\mu_i\geq 0$ and vectors $\widetilde{\mathbf{X}}_{i}=(0,\ldots,0,\tilde x_i,0,\ldots,0,\tilde y_i,0,\ldots,0)$, $i=1,\ldots,9$, such that 
\begin{equation}
\mathbf{X} = \mathbf{X}_0 + \sum_i \mu_i \widetilde{\mathbf{X}}_{i}.
\end{equation}

\section{II. Explicit derivation of the three-point correlations}

Here, we provide a detailed derivation of Eq. \eqref{eq:3point} following the general scheme presented in Ref.~\cite{Asadian}.  At the beginning of the Ramsey sequence the qubit is initialized in state $|\downarrow\rangle$. The pulse sequence $U_{RM}(\tau)$ acts on the composite qubit-resonator system. Here the $R_{\pi/2}(\phi)$ denote $\pi/2$-rotations of the qubit with an adjustable phase $\phi$, which in the basis $\{ |\downarrow\rangle, |\uparrow\rangle\}$ is  defined as 
\begin{equation}
R_{\pi/2}(\phi)=\frac{1}{\sqrt{2}}\left(\begin{array}{cc}  1 & e^{i\phi} \\  -e^{-i\phi} & 1\end{array}\right).
\end{equation}
The evolution between the pulses, $U(\tau)=\Big[\proj{\downarrow}+e^{i\phi(\tau)}\mathcal{D}_b(\alpha(\tau))\proj{\uparrow}\Big]U_0(\tau)$, describes the qubit-resonator interaction for an interaction
period $\tau$.  If at the initial time $t_0$ the resonator is in an arbitrary state $\rho_0$, the total system density operator after the pulse sequence is $U_M(\varphi, t_1) \proj{\downarrow}\otimes\rho_0 U^\dag_M(\varphi, t_1)$ and after partial trace over the qubit the oscillator's reduced density operator reads
\begin{equation}
\rho(t_1)=p_+\rho^+(t_1)+p_-\rho^-(t_1).
\end{equation}
The probabilities $p_+$ and $p_-$ for finding the qubit in state $|\uparrow\rangle$ and $|\downarrow\rangle$, respectively, are then given by 
\begin{equation} 
p_\pm(t_1) = {\rm Tr}\{ E_\pm(\varphi,\tau_1)\rho_0 E^\dagger_\pm(\varphi,\tau_1)\}.
\end{equation} 
where $E_\pm(\varphi,\tau) =\frac{1}{2}\left[\mathbbm{1} \pm e^{i  \varphi(\tau)}\mathcal{D}(\alpha(\tau)) \right]U_0(\tau)$ are Kraus operators satisfying $E_+^\dag E_++E_-^\dag E_-=\mathbbm{1}$. Right after the sequence a projective measurement of $Z$ is done on the qubit with expectation value given by
\begin{align}\nonumber
&\mean{Z_{t_1}}=p_+(t_1)-p_-(t_1)= {\rm Tr}\{ E_+(\varphi_1,\tau_1)\rho_0 E^\dagger_+(\varphi_1,\tau_1)\}
\\
&-{\rm Tr}\{ E_-(\varphi_1,\tau_1)\rho_0 E^\dagger_-(\varphi_1,\tau_1)\}
=\text{Tr}\{Q(\varphi_1, \alpha_{t_1})\rho_0\}.
\end{align}
 Depending on the measurement outcome the \emph{conditioned} resonator state is
\begin{equation} 
\rho^\pm(t_1) =   \frac{  E_\pm(\varphi_1,\tau_1)\rho_0 E^\dagger_\pm(\varphi_1,\tau_1)}{p_\pm(t_1)}.
\end{equation} 

We now generalize the above considerations for a sequence of measurements. We set $t_0=0$ the time right before the first measurement and denote by $t_n$ the time after the $n$-th RM is complete. The variables $Z(t_n)=\pm 1$ describe the outcome of the respective measurements. Each Ramsey sequence is characterized by displacement amplitudes $\alpha(\tau_n)$, an adjustable phase of the first $\pi/2$-pulse $\varphi_n$ and the geometric phases $\phi_n\equiv\phi_n(\tau_n)$.  Starting from the initial resonator density operator $\rho_0=\rho(0)$ the state conditioned on the first measurement outcome $\eta_1=\pm$ is
\begin{equation}
\begin{split}
\rho^{\eta_1}(t_1)=E_{\eta_1}&(\tau_1) \rho_0 E_{\eta_1}^\dag(\tau_1)/p_{\eta_1}=
\\\dfrac{1}{4p_\pm}\big(\rho_0(t_1)+&\mathcal{D}(\alpha_1)\rho_0(t_1)\mathcal{D}(-\alpha_1)\pm [\mathcal{D}(\alpha_1)\rho_0(t_1)+{\rm H.c.}]\big).
\end{split}
\end{equation} 
This state evolves freely for a time $t_2-\tau_2- t_1$ and then a second measurement is performed. By repeating the arguments from above, the probabilities for this second measurement, conditioned on the first outcome, are given by  
\begin{equation}
\begin{split}
p_{\pm|\eta_1}=&\\ \frac{1}{2}
\pm \frac{1}{2}{\rm Tr}&\{ÊÊÊQ(\varphi_2,\alpha(\tau_2))U_0(t_2-t_1) \rho^{\eta_1}(t_1) U_0^\dag(t_2-t_1)  \}.
\end{split}
\end{equation} 
Therefore, the conditioned expectation value of the second measurement is
\begin{equation}
\langle Z_{t_2}\rangle_{\eta_1}=p_{+|\eta_1}-p_{-|\eta_1} = {\rm Tr}\{Q(\varphi_2,\alpha(\tau_2)) \rho^{\eta_1}(t_2) \},
\end{equation}
where $\rho^{\eta_1}(t_2)=U_0(t_2-t_1) \rho_1^{\eta_1}(t_1) U_0^\dag(t_2-t_1) $ denotes the time evolved conditioned density operator.
The two point correlation function between two successive measurements is 
\begin{equation}
\begin{split}
&\langle Z_{t_2}Z_{t_1}\rangle = (p_{+|+}-p_{-|+})p_++ (p_{-|-}-p_{+|-})p_-\\
&=p_+\mean{Q(\varphi_2,\alpha(\tau_2))}_+-p_-\mean{Q(\varphi_2,\alpha(\tau_2))}_- \\
&= \text{Tr}\{Q(\varphi_2,\alpha(\tau_2))
[U_0(t_2-t_1)e^{i\varphi_1}\mathcal{D}(\alpha(\tau_1))\rho(t_1)U^\dag_0(t_2-t_1)\\
&+{\rm H.c.}]\} \nonumber\\
&=\text{Tr}\{Q(\varphi_1,\alpha_{t_1}) Q(\varphi_2,\alpha_{t_2})\rho_0\},
\end{split}
\end{equation}
where we assumed $[\mathcal{D}(\alpha_{t_1}), \mathcal{D}(\alpha_{t_2})]=0$. For a 3-point correlation function we have

\begin{align}
\langle Z_{t_3} Z_{t_2}Z_{t_1}\rangle= p_+(t_1)\langle Z_{t_3}Z_{t_2}\rangle_+- p_-(t_1)\langle Z_{t_3}Z_{t_2}\rangle_-.
\end{align}
 Simply using the above argument we have $
\langle Z_{t_3}Z_{t_2}\rangle_{\pm}=\text{Tr}\{Q(\varphi_2,\alpha_{t_2}) Q(\varphi_3,\alpha_{t_3})\rho^{\pm}(0)\}$, where $\rho^{\pm}(0)=U^\dag_0(t_1)\rho^{\pm}(t_1)U_0(t_1)$. Therefore, we have
\begin{equation}
\begin{split}
&\langle Z_{t_3} Z_{t_2}Z_{t_1}\rangle\\
&=\dfrac{1}{2}\text{Tr} \{Q(\varphi_1,\alpha_{t_1}) Q(\varphi_2,\alpha_{t_2})[U_0^\dagger(t_1)e^{i\varphi_1}\mathcal{D}(\alpha_1)\rho_0(t_1)U_0(t_1)\\
&+{\rm H.c.}]\} \\
&=\dfrac{1}{2}\text{Tr} \{Q(\varphi_1,\alpha_{t_1}) Q(\varphi_2,\alpha_{t_2})[e^{i\varphi_1}\mathcal{D}(\alpha_{t_1})\rho_0+{\rm H.c.}]\}.
\end{split}
\end{equation}
For mutually commuting displacements $\mathcal{D}(\alpha_{t_1})$, $\mathcal{D}(\alpha_{t_2})$ and $\mathcal{D}(\alpha_{t_3})$ then, we obtain
\begin{equation}
\langle Z_{t_3} Z_{t_2}Z_{t_1}\rangle=\text{Tr} \{Q(\varphi_1,\alpha_{t_1}) Q(\varphi_2,\alpha_{t_2}) Q(\varphi_2,\alpha_{t_3})\rho_0\}.
\end{equation}
This complete the derivation of Eq. \eqref{eq:3point}.

As a concrete example, let us work out the implementation of $C_3$ explicitly. For this, we need to fix $\theta=\pi/4$ in  $b=\cos(\theta)a_1+\sin(\theta) a_2$ throughout the sequence. It suffices  to consider $\tau_j\equiv\tau$ for each Ramsey measurement inducing the same required length for the displacement amplitude. Therefore, in the first measurement $\alpha_{t_1}=\alpha(\tau)e^{i\omega t_1}$. The desired amplitude is 
$\alpha_{t_1}=\sqrt{2}\alpha$ so that $\mathcal{D}_b(\alpha_{t_1})=\mathcal{D}_{a_1}
(\alpha)\mathcal{D}_{a_2}(\alpha)$. The only thing we need to fix is $|\alpha|=1.34$.
The second measurement occurs at time $t_2$ giving $\alpha_{t_2}=\alpha(\tau)e^{i\omega t_2}$, 
such that $\omega (t_2-t_1)=2\pi/3$. Therefore, we have 
$\mathcal{D}_b(\alpha_{t_2})=\mathcal{D}_{a_1}(\alpha e^{i2\pi/3})\mathcal{D}_{a_2}(\alpha e^{i2\pi/3})$. In the third measurement, $\alpha_{t_3}=\alpha(\tau)e^{i\omega t_3}$. Likewise,
we chose $\omega(t_3-t_2)=2\pi/3$ so that $\mathcal{D}_b(\alpha_{t_3})=\mathcal{D}_{a_1}(\alpha e^{-i2\pi/3})\mathcal{D}_{a_2}(\alpha e^{-i2\pi/3})$. With this, we could implement the required displacement operators appeared in $C_3$ using the sequence of Ramsey measurements. Generally, different three-point correlations of the associated modular variables composing  \eqref{eq:R} are obtained by suitable choice of $\varphi_1$, $\varphi_2$ and $\varphi_3$ in each sequence of Ramsey measurements. Similar considerations are used for implementing other parts of the PM square.

\section{III. Forced quantum harmonic oscillator}
In this section we briefly summarize the mechanisms for creating large enough spin-dependent displacements of vibrational modes as it is needed for the contextuality tests proposed in this work. The basic interaction Hamiltonian as given in Eq.~\eqref{eq:Htot} can be realized in a wide variety of systems, ranging from trapped ions to nanomechanical resonators.  For example, in the case of a trapped ion the hyperfine levels are coupled to the vibrational modes of the ion via state dependent optical dipole forces, which can be modulated at different frequencies to excite selective modes \cite{Monroe96, Monroe05}.  Equivalent state dependent forces can be implemented by coupling the motion of a trapped ion or a nanomechanical resonator to a spin 1/2 system via strong magnetic field gradients~\cite{MintertSupp, LakeSupp, Rabl09Supp}.  In this case a time-dependent coupling can be generated by simply flipping the spin state periodically. The common aspect is a modulated-in-time coupling strength which can resonantly drive the harmonic oscillator to an amplitude which is much larger than what can be achieved with a static coupling. To illustrate this we consider the Hamiltonian 
\begin{equation}
\label{eq:Hforced}
H=\omega b^\dag b+\lambda(t) b^\dag+\lambda^*(t)b,
\end{equation}
ignoring the state-dependence of the force for now. 
%
%
The time evolution in the interaction picture generated by this Hamiltonian for duration  $\tau=t-t_0$  is obtained by solving
\begin{equation}\label{eq:dotUtilde}
\partial_t \tilde U(t)= -i [\lambda(t)b^\dag e^{i\omega t}+\lambda^*(t)b e^{-i\omega t}]\tilde U(t).
\end{equation} 
We characterize the solution by the Ansatz $\tilde U(t)=e^{i\phi(\tau)}  \mathcal{D}(\tilde \alpha(\tau))$
with $\tilde{\alpha}(\tau)$ defined as
\begin{equation}
\tilde{\alpha}(\tau)=-i\int_{t_0}^t\lambda(t')e^{i\omega t'}dt',
\end{equation}
and
\begin{equation}
\phi(\tau)= \int_{t_0}^t dt' \int_{t_0}^{t'} dt''\, \lambda(t')\lambda(t'')\sin(\omega(t'-t'')).
\end{equation}
Therefore, the time evolution in Schr\"odinger picture is 
\begin{equation}
\label{eq:Utau}
U(\tau)=e^{-iH\tau} e^{i\phi(\tau)}e^{\tilde{\alpha}(\tau)a^\dag -\tilde{\alpha}^*(\tau)a},
\end{equation}
Alternatively, we can rewrite \eqref{eq:Utau} as
\begin{equation}
\label{eq:Utau1}
U(\tau)=e^{i\phi(\tau)}e^{\alpha(\tau)a^\dag -\alpha^*(\tau)a} e^{-iH\tau},
\end{equation}
in which case $\alpha(\tau)=e^{-i\omega\tau}\tilde{\alpha}(\tau)$.

\begin{figure}[!]
\centering
\includegraphics[width=1\linewidth]{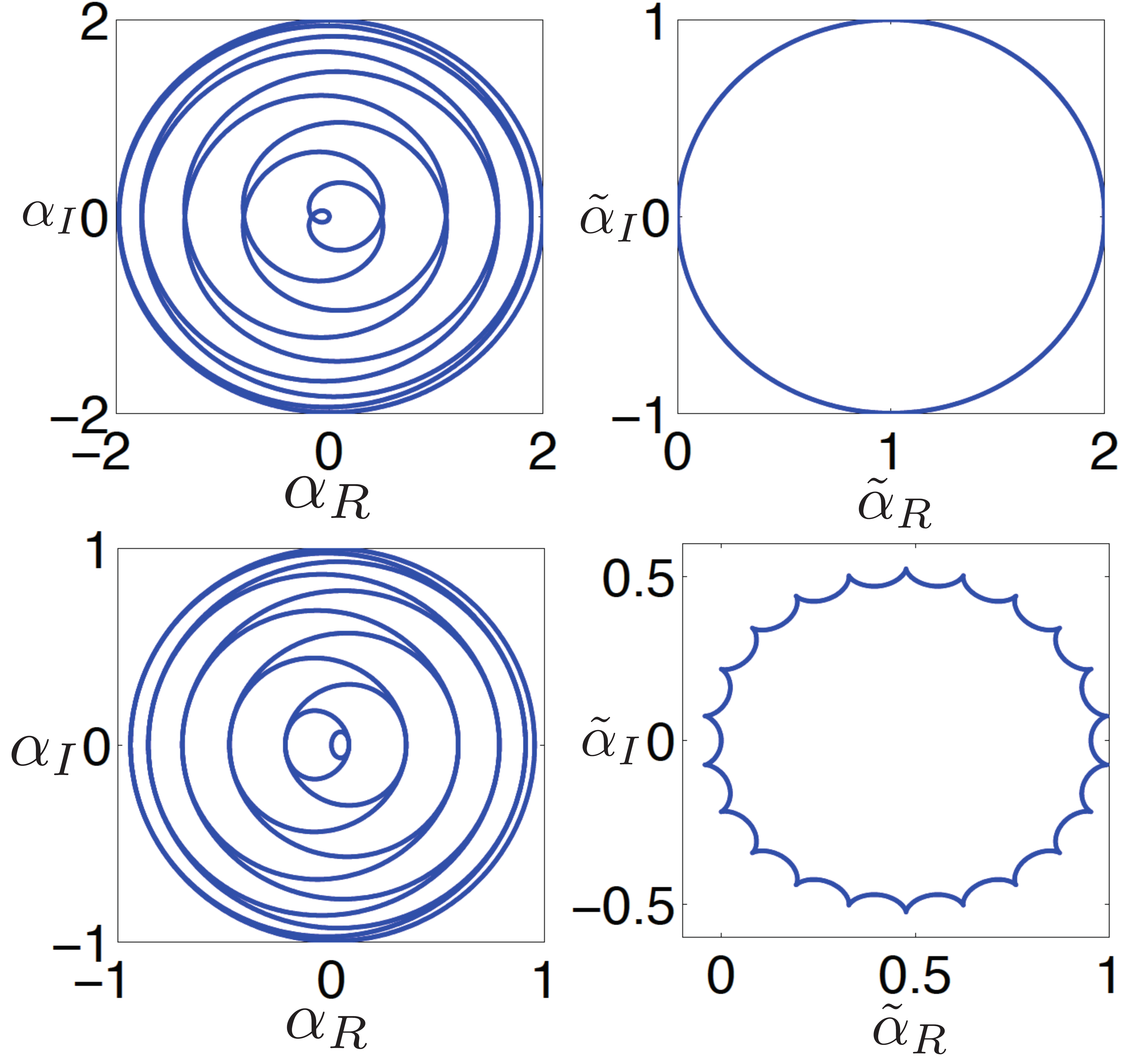}
\caption{Evolution of the displacement amplitudes $\alpha(\tau)$ (left) and $\tilde \alpha(\tau)$ (right) when excited with a near resonantly oscillating force. The upper panels correspond to the case where  $
\lambda(t)=\lambda_0 e^{-i(\omega-\delta)t}$. The lower panels correspond to the case where $\lambda(t)=\lambda_0\cos[(\omega+\delta)t]$. }
\label{ForcedPhase}
\end{figure}

Let us give examples of two different types of coupling by which large enough displacement of a vibrational mode can be achieved. One example is $
\lambda(t)=\lambda_0 e^{-i(\omega-\delta)t}$,
where $\omega$ is the natural frequency of an oscillator and $\delta=\omega_0-\omega$ characterizes the detuning of the driving (laser) field frequency from the vibrational frequency.  Therefore, we have
\begin{equation}
\label{alphat}
\begin{split}
\tilde{\alpha}(\tau)=-i\int_{t_0}^t \lambda_0 e^{i\delta t'} dt'=\dfrac{\lambda_0}{\delta}(1-e^{i\delta \tau}),\\
\phi(\tau)=\dfrac{\lambda^2_0}{\delta^2}(\delta t-\sin\delta t).
\end{split}
\end{equation}

Let us consider another type of interaction  characterized by a real coupling strength $\lambda(t)=\lambda_0\cos[(\omega+\delta)t]$, where $\omega$ is the natural frequency and $\delta=\omega_0-\omega$ is the detuning. We have
\begin{equation}
\tilde{\alpha}(\tau)=-i\int_{t_0}^t \lambda_0\cos[(\omega+\delta)t']e^{i\omega t'}dt'.
\end{equation}
This finally  (with defining $
\alpha(\tau)=\alpha_{R}(\tau)+i\alpha_{I}(\tau)$) yields
\begin{equation}
\alpha_{R}(\tau)=\dfrac{-\lambda\omega}{\delta(\delta/2+\omega)}\sin(\tau\dfrac{\delta}{2})\sin(\tau(\dfrac{\delta}{2}+\omega))
\end{equation}
and 
\begin{equation}
\alpha_{I}(\tau)=\dfrac{-\lambda\omega}{\delta(\delta+2\omega)}\left[(\delta+\omega)\sin(\tau(\delta+\omega))-
\omega\sin(\omega\tau)\right].
\end{equation}
The corresponding displacement amplitudes are shown in Fig. \ref{ForcedPhase}.


\begin{thebibliography}{99}

\bibitem{Specker60}
 E. P. Specker,
 \href{http://onlinelibrary.wiley.com/doi/10.1111/j.1746-8361.1960.tb00422.x/abstract}{Dialectica \textbf{14}, 239 (1960).}
 English translation: \href{http://arxiv.org/abs/1103.4537}{\eprint{arXiv:1103.4537}.}

\bibitem{Kochen}  S. Kochen,  and E. P. Specker,  J. Math. Mech. \textbf{17}, 59 (1967).

\bibitem{KCBS08}
 A. A. Klyachko, M. A. Can, S. Binicio\u{g}lu, and A. S. Shumovsky,
\href{http://dx.doi.org/10.1103/PhysRevLett.101.020403}{Phys. Rev. Lett. \textbf{101}, 020403 (2008).}

\bibitem{Bell64}
 J. S. Bell,
 Physics (Long Island City, N.Y.) \textbf{1}, 195 (1964).

\bibitem{Cabello08}
 A. Cabello,
 \href{http://dx.doi.org/10.1103/PhysRevLett.101.210401}{Phys. Rev. Lett. \textbf{101}, 210401 (2008).}

\bibitem{BBCP09}
 P. Badzi\c{a}g, I. Bengtsson, A. Cabello, and I. Pitowsky,
 \href{http://dx.doi.org/10.1103/PhysRevLett.103.050401}{Phys. Rev. Lett. \textbf{103}, 050401 (2009).}
 
 \bibitem{YO2012}
 S. Yu and C. H. Oh,
\href{http://dx.doi.org/10.1103/PhysRevLett.108.030402}{Phys. Rev. Lett. \textbf{108}, 030402 (2012).}

\bibitem{KBLGC12}
 M. Kleinmann, C. Budroni, J.-\AA. Larsson, O. G\"uhne, and A. Cabello,
 \href{http://dx.doi.org/10.1103/PhysRevLett.109.250402}{Phys. Rev. Lett. \textbf{109}, 250402 (2012).}

\bibitem{RH14}
 R. Ramanathan and P. Horodecki,
 \href{http://dx.doi.org/10.1103/PhysRevLett.112.040404}{Phys. Rev. Lett. \textbf{112}, 040404 (2014).}

\bibitem{CKB15}
A. Cabello, M. Kleinmann, and C. Budroni, \href{http://arxiv.org/abs/1501.03432}{arXiv:1501.03432.}

\bibitem{Roos09} G. Kirchmair, F. Z\"{a}hringer, R. Gerritsma, M. Kleinmann, O. G\"{u}hne, A. Cabello, R. Blatt, and C. F. Roos, \href{http://www.nature.com/nature/journal/v460/n7254/full/nature08172.html}{Nature (London) \textbf{460}, 494 (2009).}

\bibitem{NMR} O. Moussa, C. A. Ryan, D. G. Cory, and  R. Laflamme, \href{http://dx.doi.org/10.1103/PhysRevLett.104.160501}{Phys. Rev. Lett. \textbf{104}, 160501 (2010).}

\bibitem{Photons} E. Amselem, M. R{\aa }dmark, M. Bourennane, and A. Cabello,
 \href{http://dx.doi.org/10.1103/PhysRevLett.103.160405}{Phys. Rev. Lett. \textbf{103}, 160405 (2009).}
 
\bibitem{SciRep} J. Ahrens, E. Amselem, A. Cabello, and M. Bourennane,
\href{http://dx.doi.org/10.1038/srep02170}{Sci. Rep.\textbf{ 3}, 2170 (2013)}.

\bibitem{Zeilinger} R. Lapkiewicz, P. Li, C. Schaeff, N. Langford,
S. Ramelow, M. Wie´sniak, and A. Zeilinger, \href{http://dx.doi.org/10.1038/nature10119}{Nature (London)
\textbf{474}, 490 (2011)}.

\bibitem{Zhang2013}
X. Zhang, M. Um, J. Zhang, S. An, Y. Wang, D. Deng, C. Shen, L.-M. Duan, and K. Kim, \href{http://dx.doi.org/10.1103/PhysRevLett.110.070401}{Phys. Rev. Lett. {\bf 110}, 070401 (2013).}

\bibitem{HWVE14}
 M. Howard, J. J. Wallman, V. Veitch, and J. Emerson,
 \href{http://www.nature.com/nature/journal/v510/n7505/full/nature13460.html}{Nature (London) \textbf{510}, 351 (2014).}

\bibitem{Raussendorf13}
 R.~Raussendorf,
 \href{http://dx.doi.org/10.1103/PhysRevA.88.022322}{Phys. Rev. A \textbf{88}, 022322 (2013)}.

\bibitem{DGBR14}
 N. Delfosse, P. A. Guerin, J. Bian, and R. Raussendorf,
 \href{http://arxiv.org/abs/1409.5170}{\eprint{arXiv:1409.5170}.}

\bibitem{HHHHPB10}
 K. Horodecki, M. Horodecki, P. Horodecki, R. Horodecki, M. Paw{\l}owski, and M. Bourennane,
 \href{http://arxiv.org/abs/1006.0468}{\eprint{arXiv:1006.0468}.}

\bibitem{GBCKL14}
 O. G\"uhne, C. Budroni, A. Cabello, M. Kleinmann, and J.-\AA. Larsson,
 \href{http://dx.doi.org/10.1103/PhysRevA.89.062107}{Phys. Rev. A \textbf{89}, 062107 (2014).}

\bibitem{CAEGCXL14}
 G. Ca\~{n}as, M. Arias, S. Etcheverry, E. G\'omez, A. Cabello, G. B. Xavier, and G. Lima,
 \href{http://dx.doi.org/10.1103/PhysRevLett.113.090404}{Phys. Rev. Lett. \textbf{113}, 090404 (2014).}



\bibitem{Parity}  K. Banaszek and K. W\'odkiewicz, \href{http://dx.doi.org/10.1103/PhysRevA.58.4345}{Phys. Rev. A \textbf{58},
4345 (1998)}.

\bibitem{PlastinoPRA2010} A. R. Plastino and A. Cabello, \href{http://dx.doi.org/10.1103/PhysRevA.82.022114}{Phys. Rev. A {\bf 82}, 022114 (2010)}.

\bibitem{Massar}
 S. Massar and S. Pironio,
 \href{http://dx.doi.org/10.1103/PhysRevA.64.062108}{Phys. Rev. A {\bf 64}, 062108 (2001)}.

\bibitem{Asadian} A. Asadian, C. Brukner, P. Rabl, \href{http://dx.doi.org/10.1103/PhysRevLett.112.190402}{Phys. Rev. Lett. \textbf{112}, 190402 (2014)}.

\bibitem{Hornberger} C. Gneiting and K. Hornberger, \href{http://dx.doi.org/10.1103/PhysRevLett.106.210501}{Phys. Rev. Lett. \textbf{106}, 210501
(2011).}

\bibitem{Zhang} C. Zhang, S. Yu, Q. Chen, C.H. Oh, \href{http://dx.doi.org/10.1103/PhysRevLett.111.190501}{Phys. Rev. Lett. \textbf{111}, 190501 (2013).}




\bibitem{Peres90}
 A.~Peres,
  \href{http://www.sciencedirect.com/science/article/pii/037596019090172K}{Phys. Lett. A \textbf{151}, 107 (1990).}

\bibitem{Mermin90}
 N.D.~Mermin,
\href{http://dx.doi.org/10.1103/PhysRevLett.65.3373}
{Phys.~Rev.~Lett. \textbf{65}, 3373 (1990)}.



\bibitem{Guhne2010}
O. G\"uhne, M. Kleinmann, A. Cabello, J.-\AA ~ Larsson, G. Kirchmair, F. Z\"ahringer, R. Gerritsma, and C. F. Roos, \href{http://dx.doi.org/10.1103/PhysRevA.81.022121}
{Phys. Rev. A {\bf 81}, 022121 (2010)}.

\bibitem{Guhne13} J. Szangolies, M. Kleinmann, and O. G\"uhne, \href{http://dx.doi.org/10.1103/PhysRevA.87.050101}{Phys. Rev. A 87, 050101(R) (2013).}

\bibitem{Vourdas2004} A Vourdas, \href{http://dx.doi.org/10.1088/0034-4885/67/3/R03}{Rep. Prog. Phys. {\bf 67}, 267 (2004)}.

\bibitem{Zhang2006} S.-Z. Zhang, X.-T. Xie, and W.-X. Yang, Communications in Theoretical Physics {\bf 46}, 306 (2006).


\bibitem{Monroe96} C. Monroe, D. Meekhof, B. King, D. Wineland, \href{ 10.1126/science.272.5265.1131 }{Science \textbf{272}, 1131 (1996).}

\bibitem{Monroe05} P. C. Haljan, K.-A. Brickman, L. Deslauriers, P. J. Lee, and C. Monroe, 
\href{http://dx.doi.org/10.1103/PhysRevLett.94.153602}{Phys. Rev. Lett. \textbf{94}, 153602 (2005).}




\bibitem{MintertSupp} F. Mintert and C. Wunderlich, \href{http://dx.doi.org/10.1103/PhysRevLett.87.257904}{Phys. Rev. Lett. \textbf{87}, 257904
(2001).}


\bibitem{LakeSupp} K. Lake, S. Weidt, J. Randall, E. D. Standing, S. C. Webster, and W. K. Hensinger, \href{http://dx.doi.org/10.1103/PhysRevA.91.012319}{Phys. Rev. A \textbf{91}, 012319 (2015).}

\bibitem{Rabl09Supp}P. Rabl, P. Cappellaro, M. V. Gurudev Dutt, L. Jiang, J. R. Maze, and M. D. Lukin, \href{http://dx.doi.org/10.1103/PhysRevB.79.041302}{Phys. Rev. B \textbf{79}, 041302(R) (2009).}



\end{thebibliography}
\end{document}